\documentclass[preprint]{IEEEtran}
\usepackage{mathtools}
\usepackage{graphicx}
\usepackage{amssymb}
\usepackage{bm}
\usepackage{cleveref}
\usepackage{subcaption}
\usepackage{cite}
\usepackage{flushend}

\usepackage{amsthm}
\usepackage{color}

\allowdisplaybreaks

\newcommand{\x}{\mathbf{x}}

\newcommand{\norm}[1]{\left\lVert#1\right\rVert}

\DeclareMathOperator*{\argmin}{arg\,min}

\title{AutoEKF: Scalable System Identification for COVID-19 Forecasting from Large-Scale GPS Data}

\author{Francisco Barreras$^*$ \qquad Mikhail Hayhoe$^\dagger$ \qquad Hamed Hassani$^\dagger$ \qquad Victor M. Preciado$^\dagger$
\thanks{$^*$Department of Mathematics, University of Pennsylvania, \texttt{fbarrer@sas.upenn.edu}
\newline\indent$^\dagger$Department of Electrical \& Systems Engineering, University of Pennsylvania, \texttt{\{mhayhoe, hassani, preciado\}@seas.upenn.edu}
\newline\indent We thank Duncan Watts, Mark Whiting, Homa Hosseinmardi, and Amir Ghassemian for their helpful comments and insightful discussions. This work was supported, in part, by the National Science Foundation under awards NSF-TRIPODS-1934876, CAREER-ECCS-1651433, NSF-III-2008456, the Rockefeller Foundation, and the Amazon COVID-19 HPC Consortium.
}}

\begin{document}
\maketitle
\begin{abstract}
We present an Extended Kalman Filter framework for system identification and control of a stochastic high-dimensional epidemic model. The scale and severity of the COVID-19 emergency have highlighted the need for accurate forecasts of the state of the pandemic at a high resolution. Mechanistic compartmental models are widely used to produce such forecasts and assist in the design of control and relief policies. Unfortunately, the scale and stochastic nature of many of these models often makes the estimation of their parameters difficult. With the goal of calibrating a high dimensional COVID-19 model using low-level mobility data, we introduce a method for tractable maximum likelihood estimation that combines tools from Bayesian inference with scalable optimization techniques from machine learning. The proposed approach uses automatic-backward differentiation to directly compute the gradient of the likelihood of COVID-19 incidence and death data. The likelihood of the observations is estimated recursively using an Extended Kalman Filter and can be easily optimized using gradient-based methods to compute maximum likelihood estimators. Our compartmental model is trained using GPS mobility data that measures the mobility patterns of millions of mobile phones across the United States. We show that, after calibrating against incidence and deaths data from the city of Philadelphia, our model is able to produce an accurate 30-day forecast of the evolution of the pandemic.
\end{abstract}
\maketitle

\section{Introduction}

The ongoing COVID-19 pandemic has highlighted the need for data-driven epidemic models capturing low-level human interaction as well as rapidly changing mobility patterns. Accurately and robustly predicting the evolution of an epidemic is important for evaluating the impact of policies such as social distancing~\cite{chang2021mobility}, travel restrictions~\cite{bajardi2011human}, and vaccine distribution~\cite{preciado2014optimal}. At the same time, predictions of inherently stochastic dynamics using noisy observations carry significant uncertainty that needs to be understood for the design of such policies. Historically, the data required to analyze high-resolution interaction patterns in the population have been unavailable due to technological and privacy concerns. Recently, multiple datasets describing human mobility and behavior have been made available, signifying a revolution in the data-driven analysis of epidemic models. Models that consider the population structure at a higher granularity can, theoretically, replicate important features of real-world epidemics such as resurgence and high variation in the final fraction of infected individuals \cite{watts2005multiscale}, which coarser models often fail to exhibit. 

Epidemic models can be broadly segregated into two frameworks: model-free machine learning methods, and mechanistic models derived from first principles. Machine learning model-free frameworks have been shown to be capable of accurately forecasting epidemic trajectories with few assumptions \cite{rodriguez2020deepcovid, bhouri2020covid} using efficient gradient methods. Unfortunately, these methods require large amounts of training data, lack interpretability, and quantifying and understanding uncertainty in their predictions is difficult. On the other hand, mechanistic models characterize epidemic dynamics from first principles, are easily interpretable, can incorporate constraints and can directly model stochasticity \cite{nowzari2016analysis}. However, despite their simple formulation and realism, the stochasticity and nonlinearity in some epidemic models makes the task of identifying their parameters and initial conditions difficult, since the error in forecasting typically grows when integrating into the future. Moreover, deriving the exact distribution of the state of the system is typically intractable \cite{arulampalam2002tutorial}.

In the context of epidemics, fine-grained mechanistic models are typically high-dimensional, with thousands of latent states associated with different subpopulations. The initial conditions of these models can be regarded as additional unknown parameters and, thus, one must identify a large number of unknown parameters using little data usually in the form of aggregated incidence and death counts. A common strategy is finding parameters that minimize a loss function using zero order methods \cite{chang2021mobility,balcan2009multiscale, balcan2010modeling, chinazzi2020effect, ferguson2020report, lorch2020quantifying, davis2020estimating}. Due to the cascading error, estimators based on repeated sampling of trajectories require a large number of simulations to reduce their variance. Additionally, zero order methods are known to scale poorly compared to first order methods.  As a consequence, models fit using this strategy often set parameters to values from clinical literature or values guided by heuristics. This is often the case for the initial conditions of the latent states, usually assumed to be a small number of ``seeds'' sampled uniformly at random \cite{chang2021mobility, lorch2020quantifying, li2020substantial, pei2020differential, pei2020initial}. These heuristics can lead to misspecification of other parameters, and are also impractical when calibrating a model initialized at any point other than the beginning of the epidemic, when observations are less reliable.
 
Data-assimilation methods present an alternative for inference and system identification that handles growth of error by repeatedly adjusting the forecasts based on observed data. These Bayesian updates reduce the uncertainty of the estimates by filtering out trajectories that disagree with the observations \cite{arulampalam2002tutorial}. One such method, the extended Kalman filter (EKF) \cite{daum2005nonlinear}, uses simple recursive update equations to generate a posterior estimate of the model state. Amongst its many qualities, the EKF allows for recursive computation of the derivatives of the data likelihood with respect to the parameters, the so-called sensitivity derivatives \cite{astrom1979maximum}, which can be used for system identification with gradient-based methods.

Traditionally, the EKF has been regarded as intractable for high-dimensional models due to the cost of maintaining and operating on the state covariance matrix and, thus, it is seldom used in fitting epidemic models. Instead, other techniques based on particle filters are typically used \cite{ionides2006inference,  shaman2012forecasting, li2020substantial, pei2020initial, pei2020differential}. However, the estimation algorithms involved are often sensitive to hyperparameters, their convergence guarantees rely on restrictive assumptions, and their stochastic estimates of the log-likelihood gradient might require a high number of iterations and particles to be reliable \cite{ionides2006inference, lindstrom2012efficient}.

In this paper, we introduce AutoEKF, a hybrid approach between data-driven and mechanistic models, which combines a high-dimensional fine-grained meta-population model with a scalable and highly parallel implementation of the EKF for inference and forecasting of COVID-19. By using two staple tools from deep learning, automatic differentiation \cite{baydin2018automatic} and massive parallelization of linear algebra sub-routines, AutoEKF can efficiently carry out the EKF Bayesian updates and find the gradient of the data-likelihood. This gradient is used to estimate the maximum likelihood parameters and initial conditions using gradient-based methods for a model with thousands of states. We remark that we can compute the likelihood's gradient exactly and efficiently, unlike for particle filter methods which only approximate it.

We use human mobility data to construct high-resolution dynamical contact networks between individuals as part of our proposed networked meta-population model. Integrating such data into epidemic models commonly involves constructing a network over which the epidemic spreads, with nodes representing groups of individuals and edges representing interaction patterns between them. Such network representations range in scale and resolution, from long-range mobility patterns (i.e. trains, flights, highways) \cite{guimera2005worldwide, wu2006transport}, to local commuting patterns \cite{de2007structure, li2020substantial}, down to the level of contacts at places of interest \cite{chang2021mobility, birge2020controlling, aleta2020modelling}. To construct these networks, we follow a methodology similar to \cite{chang2021mobility}; however, we treat the sparsity in the data more carefully by implementing a Bayesian inference approach. We use these contact networks in conjunction with AutoEKF to calibrate an epidemic model for the city of Philadelphia against reported cases and deaths of COVID-19 gathered from public sources. Our approach is fully data-driven in that all of our parameters and initial conditions are learnt directly from the data.

\section{Background}

We study the problem of calibrating a high-dimensional mechanistic epidemic model for inference and forecasting of COVID-19 using fine-grained human mobility data. This involves constructing fine-grained contact networks from sparse visitation data, which requires inference about the distribution of unobserved visits. In system identification, we jointly estimate the latent states of the model as well as identify unknown parameters and initial conditions using noisy aggregated data in the form of daily reported cases and deaths of COVID-19. 

\subsection{Datasets}

The primary datasets used in this work are case counts and deaths due to COVID-19, as well as mobility data aggregated from GPS-enabled device locations. Data on the number of daily cases and deaths due to COVID-19 is taken from The New York Times~\cite{nytimes}. Mobility data is collected by SafeGraph~\cite{safegraph}, which collates anonymized GPS-enabled device location data from thousands of mobile apps in the form of hourly visits to places of interest (POIs) such as grocery stores, restaurants, and airports, among others. This data includes roughly $5\%$ of all GPS-enabled devices in the U.S. and, thus, is inherently sparse with an unknown sampling bias. The data includes aggregated information on the originating destination of visits, i.e., the census tract (CT) where an individual resides, as well as the median visit duration and approximate area (in square feet) of each POI. For privacy reasons, SafeGraph separately presents aggregate weekly counts for visitors' originating destinations and the hourly visits to the POI. Hence, there is no information on the temporal distribution of visits attributed to a specific census tract.

\subsection{Epidemic Model} \label{sec:model}
The epidemic model in this work follows the standard SEIR stochastic model with a deceased compartment \cite{nowzari2016analysis} between multiple populations. In particular, we treat each census tract as a separate subpopulation in a stochastic metapopulation model and study the interactions between them. Census tract $i$ at day $t$ has its population divided into compartments $S_i^{(t)}, E_i^{(t)}, I_i^{(t)}, R_i^{(t)}$ and $D_i^{(t)}$ corresponding to susceptible, exposed, infected, recovered, and deceased individuals, respectively. To study the inter-census-tract interactions, we construct daily contact networks from mobility data (described in detail in \Cref{sec:networks}) which are used to estimate the expected number of contacts between census tracts. The stochastic transitions between compartments at day $t$ are given by 
\begingroup
\addtolength{\jot}{0.75em}
\begin{align*}
\begin{split}
    \left(S_i \to E_i\right)^{(t)} &\sim \operatorname{Poisson}\left(\beta \sum_{j \in \mathcal{C}} M_{i,j}^{(t)} \dfrac{S_i^{(t)} I_j^{(t)}}{N_i N_j}\right)  \\[-5pt]
    \left(E_i \to I_i\right)^{(t)} &\sim \operatorname{Poisson}\left(\kappa E_{i}^{(t)}\right)\\[-5pt]
    \left(I_i \to R_i \right)^{(t)} &\sim \operatorname{Poisson}\left(\delta I_{i}^{(t)}\right) \\[-5pt]
    \left(I_i \to D_i \right)^{(t)} &\sim \operatorname{Poisson}\left(\rho I_{i}^{(t)}\right) 
\end{split}
\end{align*}
\endgroup
where $\mathcal{C}$ is the set of census tracts, $M_{ij}^{(t)}$ is the expected number of contacts between CTs $i$ and $j$ on day $t$, $\beta$ is the rate of infection from a single contact between susceptible and infected individuals, $\kappa$ is the inverse of the latency period, i.e., the rate of transition from exposed to infected, $\delta$ is the recovery rate for infected individuals, $\rho$ is the fatality rate, and $N_i$ is the population of CT $i$. This transition model is closely related to previous work \cite{chang2021mobility, li2020substantial, chinazzi2020effect} so we spare details for simplicity.

\section{Contact networks from GPS mobility data} \label{sec:networks}

In this work we construct high-resolution dynamic contact networks between census tracts (CTs) of the region of analysis, whose edges are estimates of the expected number of contacts between individuals from two census tracts in a given hour. We construct these networks using GPS-enabled device location data provided by SafeGraph~\cite{safegraph}.

In the model, transmissions of the virus occur when two individuals are in the same place of interest (POI) at the same time. Let $\mathcal{P}$ denote the set of POIs, $\mathcal{C}$ the set of CTs, and $y_{pi}^{(t)}$ the number of number of visits from CT $i \in \mathcal{C}$ to POI $p \in \mathcal{P}$ at hour $t$. Following \cite{chang2021mobility}, we define our matrix of contacts as $M_{ij}^{(t)} \coloneqq \delta_p^2 y_{pi}^{(t)}y_{pj}^{(t)} / A_p$, where $\delta_p$ is the median visit duration to $p$ and $A_p$ is its area. In what follows, we address the challenge of estimating $y_{pi}^{(t)}$ using highly granular, but sparse, visitation data.

SafeGraph only provides mobility patterns for a fraction of the population. Indeed, the percentage of observed individuals varies across CTs and changes from week to week; nonetheless, these sampling rates can be inferred by comparing the observed GPS-enabled device counts with the population of each CT. Furthermore, to ensure privacy, SafeGraph only provides counts for CT of origin aggregated at a weekly resolution, and use differential privacy~\cite{dwork2014algorithmic} to corrupt these counts. In contrast, the temporal distribution of visits is provided at an hourly resolution as aggregated counts across all CTs.

Let $z_{pi}$ denote the total number of visits to POI $p$ originating from CT $i$ in the whole week\footnote{Since SafeGraph only provides daily counts of visitors from all CTs and counts of weekly aggregated visits from a given CT, we assume $z_{pi}$ is proportional to the number of visitors.}, and $\theta_i$ denote the fraction of observed devices from CT $i$. The observed random variables are denoted as $v_p^{(t)}$ for the observed visits at POI $p$ for time $t$, stacked into the vector $\mathbf{v}_p$ for the whole week, and $x_{pi}$ for the aggregated count of observed visits to POI $p$ from CT $i$ for the whole week. Estimation of $y_{pi}^{(t)}$ has been addressed in previous work using iterative proportional fitting \cite{chang2021mobility} where the joint distribution of $y_{pi}^{(t)}$ is estimated from its marginals. However, the approach in \cite{chang2021mobility} relies on estimates of devices staying at home, a variable inaccurately measured due to selection bias \cite{safegraph_technical}. In contrast, we simply assume that the distribution of visits over the week and their CT of origin are independent and focus on addressing the issue of sparsity. 

The fraction of devices observed by SafeGraph in a typical week during the COVID-19 pandemic is roughly $5\%$, with values of $\theta_i$ as low as 1\% for some CTs; hence, the vast majority of visits are unobserved. Moreover, the vector of hourly visits over the week in a typical POI is highly sparse.
Clearly, estimating $y_{pi}^{(t)}$ requires inference of when these unobserved visits occurred and of their CT attribution. Correcting the sampling bias by rescaling observed visit counts will highly overestimate the concentration of visits in POIs with sparse data and, thus, highly overestimate the number of contacts.

\begin{figure}[ht!]
    \centering
    \def\svgwidth{0.99\columnwidth}
    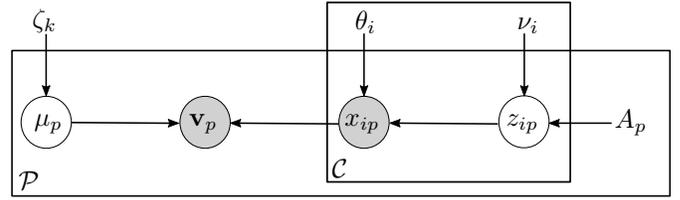
    \caption{Graphical model generating visits from each CT in $\mathcal{C}$ to each POI in $\mathcal{P}$ for a cluster $k$ of POIs for a single week.}
    \label{fig:graphical}
\end{figure}

To address this sparsity issue, we use a Bayesian inference approach in which we assume that POIs with similar visitation patterns are parameterized by the same priors, in a hierarchical model that describes the distribution of the visits. Inferring the temporal distribution of visits in a POI by using similar POIs as a proxy reduces the variance of the estimations, since it increases the sample size, and yields less sparse estimates since there is a regularization towards a non-sparse prior. To implement this approach, we make the assumption that $\mathbf{v}_{p}$ follows a multinomial distribution with parameters $\mu_p$ and $\sum_{i\in \mathcal{C}} x_{pi}$, where $\mu_p$ is drawn from a $\text{Dirichlet}(\zeta)$ prior, and represents the temporal distribution of any visit. Furthermore, we assume that $x_{pi}$ is drawn from a $\text{Binomial}(z_{pi},\theta_i)$ distribution, where we assume that $z_{pi}$ has a prior distributed as $\text{Poisson}(\nu_i A_p)$ where $\nu_i$ is an intensity parameter. Our crucial assumption is that the parameters $\zeta$ and $\nu_i$ are shared across POIs within the same cluster $k$. This generative model is summarized in Figure \ref{fig:graphical}.

In Subsection \ref{sec:clustering} we explain a methodology to cluster POIs according to their proximity and visitation patterns. With those clusters, we set the parameters $\zeta$ and $\nu_i$ for each cluster using maximum likelihood estimation. We can infer the posterior distribution of $\mu_p$ and $z_{pi}$ using standard Bayesian inference methods \cite{koller2009probabilistic}, which is facilitated by their choice of distribution as conjugate priors of $\mathbf{v}_p$ and $x_{pi}$, respectively. Our independence assumption is that $\mathbf{y}_{pi}$, the stacked vector of $y_{pi}^{(t)}$, is distributed $\text{Multinomial}(\sum_{i\in \mathcal{C}} z_{pi}, \mu_p)$, and hence, with the posterior distributions of $\mu_p$ and $z_{pi}$ we can compute the expected number of contacts between different CTs, $M_{ij}^{(t)}$, in a straightforward manner.

\subsection{Clustering points of interest} \label{sec:clustering}

We wish to cluster places of interest based on both their geographical location as well as their patterns of visitations. For this reason, we employ an approach that couples the clustering of time-series and networks, called the CCTN algorithm~\cite{liu2019coupled}. The underlying idea is to construct an embedding of the POIs, i.e., a low-dimensional representation in which ``similar'' POIs are close to one another, and then to find clusters in the latent embedding space. Since the geographical interpretation of clusters is important in our context, we perform a further clustering on any large groups of POIs to split them according to distance from one another.

In this setting we have $N$ POIs, each with a (commonly sparse) time series of daily visits per square foot over $T$ days, which we stack into the rows of the matrix $X\in\mathbb{R}^{N\times T}$. We use visits per square foot since we wish to cluster POIs with similar \emph{patterns} of visitations, not just numbers of visits. The main idea of the embedding is to reconstruct the matrix $X$ as $\tilde{X} = CW$, where $W \in \mathbb{R}^{d\times T}$ is the basis matrix, representing $d$ time series patterns learned from data, and $C \in \mathbb{R}^{N\times d}$ is the embedding matrix which reconstructs the original time series of visits for each POI using weighted combinations of the $d$ basis patterns in $W$. To incorporate the geographical locations of the POIs, we build a network wherein nodes represent POIs and edge weights are inversely proportional to geographical distance (dropping to zero, i.e., no edge, after a fixed distance). Then, the embedding is regularized for smoothness across the network: we ensure nodes' embeddings are not dramatically different than their network neighbors' embeddings. Hence, the embedding matrix $C$ and basis patterns $W$ are found by solving the problem
\begin{align}
    \argmin_{C,W} \norm{X - \tilde{X}}_F^2 + \lambda\cdot\text{tr}(C^\intercal L C),\label{eq:embedding}
\end{align}
where $L$ denotes the graph Laplacian matrix, and $\lambda$ is the parameter which controls the strength of the network regularization. In order to solve this problem, the CCTN algorithm performs an alternating process, fixing $C$ and solving a quadratic problem for $W$, then fixing $W$ and solving a mixed-integer program to find $C$. For full implementation details, see~\cite{liu2019coupled}. We then perform k-means clustering in the $d$-dimensional embedding space described by $C$ in order to cluster the POIs. This primary clustering focuses on grouping POIs with similar visitation patterns and regularizes for local smoothness in the network, but it does not necessarily disincentivize POIs which are geographically distant from being clustered together. Thus, in any clusters with many POIs, we perform a further k-means clustering using the geographical location of the POIs to split these larger clusters based solely on distance. The final result of this approach is clusters of POIs that are geographically close, and exhibit similar visitation patterns (in terms of daily visits per square foot) to those within their group.

\section{System Identification and Inference}

Our system identification method, AutoEKF, is a scalable implementation of the extended Kalman filter that uses automatic differentiation and massive parallelization to find the maximum likelihood estimate of the unknown parameters. The EKF Bayesian update equations are exceedingly simple and allow for recursive computation of the gradient of the data log-likelihood (the score function), also known as \textit{sensitivity derivatives} \cite{astrom1979maximum}. Indeed, in a nonlinear system given by
\begin{align*}
    \mathbf{x}_{t} &= f_{\theta}(\mathbf{x}_{t-1}) + \mathbf{v}_{t}, \qquad \mathbf{v}_t \sim \mathcal{N}(\mathbf{0}, \mathbf{Q}_t), \\
    \mathbf{y}_{t} &= g_{\theta}(\mathbf{x}_{t}) + \mathbf{w}_{t}, \qquad \mathbf{w}_t \sim \mathcal{N}(\mathbf{0}, \mathbf{R}_t),
\end{align*}

\noindent with latent states $\mathbf{x}_t$ and observations $\mathbf{y}_{t}$, applying the EKF yields the following expression for the data log-likelihood
\begin{align*}
    \log p_{\theta}(\mathbf{\mathbf{y}_{1:T}}) = \sum_{t=2}^T \log \mathcal{N}\left(\mathbf{y}_t ; \mathbf{J}_g \hat{\mathbf{x}}_{t|t-1}, \mathbf{J}_g \hat{\mathbf{P}}_{t|t-1} \mathbf{J}_g^\intercal\right),
\end{align*}
\noindent where $\mathbf{J}_g$ is the Jacobian of $g$, and $\hat{\mathbf{x}}_{t|t-1}$ and $\hat{\mathbf{P}}_{t|t-1}$ are the estimates of the mean and covariance of $\mathbf{x}_t$, which are updated recursively. Thus, the score function can be obtain by recursively computing $\nabla_{\theta} \hat{\mathbf{x}}_{t|t-1}$ and  $\nabla_{\theta} \hat{\mathbf{P}}_{t|t-1}$.

The EKF has traditionally been regarded as intractable for high-dimensional models due to the cost of maintaining and operating on the matrix $\hat{\mathbf{P}}_{t|t-1}$ and, thus, particle data-assimilation methods are often preferred (see \cite{ionides2006inference, li2020substantial, pei2020differential, pei2020initial, shaman2012forecasting} for examples in epidemic models). However, algorithms for parameter estimation with particle filters often rely on approximations of the score function \cite{schon2015sequential} and can be sensitive to hyperparameters, which causes issues for their convergence to local optima in practice \cite{lindstrom2012efficient}. In contrast, we propose a tractable implementation of the EKF that efficiently computes the score function to obtain the maximum likelihood estimate with off-the-shelf machine learning tools.

The efficiency of AutoEKF is achieved by means of automatic differentiation and massive parallelization of linear algebra routines, resulting in cheap evaluations of the score function when running on a GPU. We remark that this technique is also accessible, since machine learning libraries can differentiate and parallelize basic implementations of the EKF.

\section{Simulation Case Study: Philadelphia}

In order to validate our approach we calibrate a metapopulation model using AutoEKF on data from the city of Philadelphia. We construct a network representing contacts between 376 census tracts and fit to incidence and death data at the county level for the period from July 20th, 2020 to October 18th, 2020 obtained from the New York Times repository. This involves identifying the model parameters $(\beta, \kappa, \gamma, \rho)$ as well as initial conditions for each of the $1,128$ latent states corresponding to the compartments $S_i^{(0)}, E_i^{(0)}$ and $I_i^{(0)}$ for each CT $i$\footnote{The initial conditions for compartments R and D are not learned as they have no effect on the evolution of the system.}. All experiments were run on an Amazon P3 server instance with a Tesla V100 GPU and 61GB of RAM.

\subsection{Construction of Networks} \label{sec:construction_nets}

We cluster 17,803 POIs for the city of Philadelphia as described in \Cref{sec:clustering} to form 800 clusters of POIs with similar visitation patterns. These clusters were constructed using the time series of visits ranging from May 1st, 2020 to July 20th, 2020, the period preceding our experiments. We used these clusters to perform maximum likelihood estimation and posterior inference according to the data generation model discussed in \Cref{sec:networks}. We evaluate the quality of these networks by comparing with two simple benchmarks: networks constructed with no clustering (where each POI is its own cluster), and networks constructed with clustering by census tract membership of the POIs.

\begin{figure}[ht!]
    \centering
    \includegraphics[width=\columnwidth]{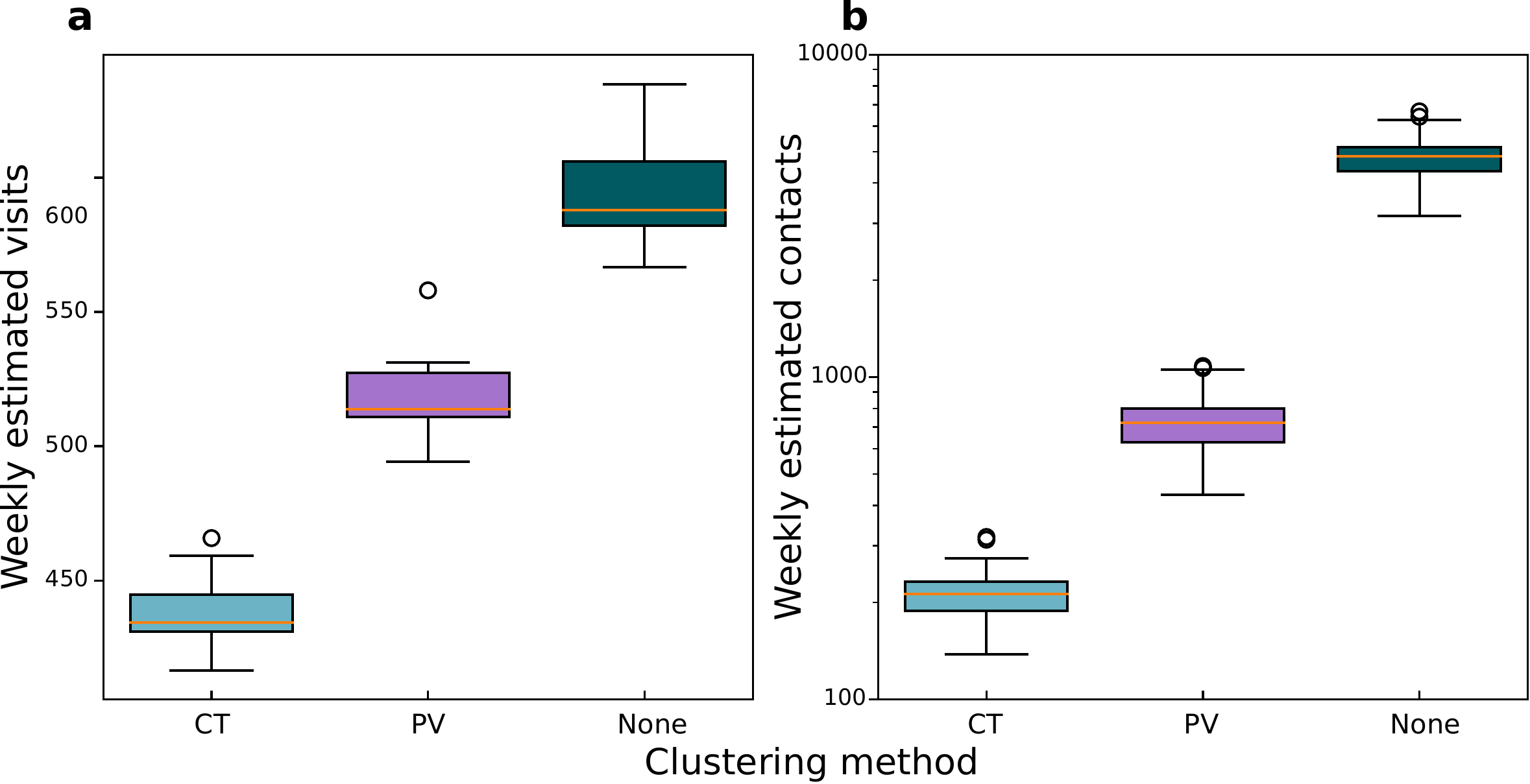}
    \caption{Choice of POI clustering has a sizeable impact on the contact networks. \textbf{a}, the estimate for the unobserved visitors to a POI from a given CT is overestimated when using sparse data and underestimated when using uninformative priors. \textbf{b}, These differences in scale increase for the number of contacts since the inferred distributions of visits are highly concentrated.}
    \label{fig:diff_networks}
\end{figure}

For each of these clustering methods, we estimated the posterior distribution of visits and contacts taking place in each POI using Bayesian inference on the model described in Figure \ref{fig:graphical}. We used maximum likelihood estimation to produce the priors $\zeta_k$ and $\{\nu_{ik}\}_{i \in \mathcal{C}}$ for each cluster $k$. In Figure \ref{fig:diff_networks} we compare the inferred number of expected contacts, aggregated by day, of the networks generated with each of these clustering choices. The choice of clustering has a sizeable impact on the resulting contact networks. Figure \ref{fig:diff_networks} a, shows that the estimate for the unobserved visitors to a POI from a given CT tends to be overestimated if only information from that POI is used. This is a consequence of the ``shrinking'' towards the prior coming from the Bayesian inference if the POIs are grouped into clusters. This effect is larger when the cluster of a given POI contains many other POIs without visitors from the same CTs, as is the case with the uninformative clustering by CT membership. Figure \ref{fig:diff_networks} shows that these differences in scale are further accentuated when estimating the number of contacts (notice the log-scale). This is a consequence of the concentrated estimate of visits distribution for a given POI, which results from the sparsity of the data in a single POI.

\subsection{Parameter estimation}

We validate the efficacy of AutoEKF by calibrating the model described in \Cref{sec:model} using our estimated contact networks against county-level incidence and death data from July 20th through October 18th, 2020. Due to the known inconsistencies and delays in reporting, we use the rolling 7-day average of case incidence and deaths and lag the data by $T_d$ days. A parameter tuned by cross-validation. 

\begin{figure*}[ht!]
    \centering
    \includegraphics[width=0.8\linewidth]{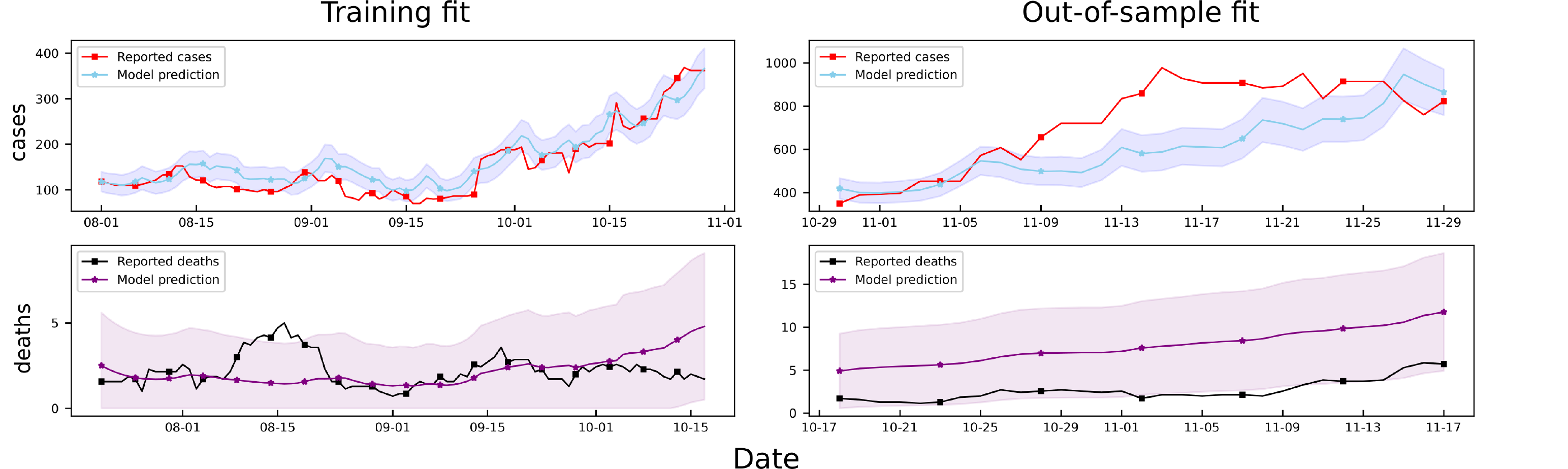}
    \caption{Left, using AutoEKF, we calibrated our model to the trajectories of cases and deaths in Philadelphia during the period from July 20th to October 18th, 2020. The means of the inferred distribution of the trajectories are represented by blue (cases) and purple (deaths) lines, respectively, and shaded regions represent symmetric 95\% CI. The inferred distribution accurately fits the data (log-likelihood = -654). Right, the 30 day forecast of the trajectories of cases and deaths for a held-out period from October 18th, 2020 to November 18th, 2020 (log-likelihood = -560). The forecasts are accurate for the first half of the testing period, and subsequently degenerate. This is expected since the error in forecast compounds when integrating into the future, highlighting the need to reinitialize the model when new data becomes available.}
    \label{fig:performance}
\end{figure*}

Specifically, our model was trained by optimizing the gradient of the data log-likelihood using the Adam optimizer \cite{adam}. The gradient of the data log-likelihood was computed using automatic differentiation on a simple implementation of the EKF and compiled for optimization in a GPU. All of our code is written in Python3 using Numpy and JAX \cite{frostig2018compiling}, a library that supports automatic differentiation of native Python and NumPy functions, as well as a high-performance compiler of linear algebra routines, which allows for massively parallel implementations of pure Numpy and Python programs. To counter the possible multimodality of the likelihood function, we used 50 different restarts for our gradient-descent optimizer and the initial values for the unknown parameters were sampled from a wide range of plausible values using Latin hypercube sampling, and the best model was selected based on performance on a test set. The trained model was validated by measuring the likelihood of its forecast on held-out set consisting of the period between October 18th, 2020 and November 18th, 2020. The training fit using AutoEKF and out-of-sample forecast are presented in Figure \ref{fig:performance}. These figures illustrate that our model is able to accurately fit observed case and death data for the city of Philadelphia in the training period and produce highly accurate forecasts. We remark that, when forecasting with our model out-of-sample, we use mobility data from the test period as input. Thus, we are evaluating the performance of the model conditional on having mobility data as input, in a time period not used for calibration. 

We investigate the robustness of these results by training models using the benchmark networks discussed in \Cref{sec:construction_nets}. The results are summarized in Table \ref{tab:performance}. We observe a significant improvement when using Bayesian inference to impute missing visits compared to the ``no-clustering'' benchmark, and further improvement when clustering POIs according to visitation patterns as opposed to simple census tract membership. This demonstrates that the input mobility data as well as the methodology for estimating contact networks can have a major impact in the performance of meta-population models and their calibration. 

\begin{table}[ht!]
    \centering
    \begin{tabular}{|l|c|c|c|c|}
        \hline
        \begin{minipage}{0.09\linewidth}Method\end{minipage} & \begin{minipage}{0.18\linewidth} Log-likelihood EKF (IS)\end{minipage} & \begin{minipage}{0.20\linewidth}Log-likelihood forecast (OOS) \end{minipage}& \begin{minipage}{0.14\linewidth} Cases RMSE (OOS) \end{minipage}&\begin{minipage}{0.14\linewidth} Deaths RMSE (OOS)\end{minipage}\\
        \hline
        None & -660.17 & -2864.52 & 387.08 & \textbf{3.0}\\
        CT & -674.07 & -1921.06 & 331.10 & 4.3\\
        PV & \textbf{-654.07} & \textbf{-560.05}& \textbf{182.93}& 5.0\\
        \hline
    \end{tabular}
    \caption{Model fit in-sample (IS) and out-of-sample (OOS) for different choices of POI clustering methods corresponding to no clustering (None), clustering by census tract (CT), and clustering by proximity and visitation patterns (PV).}
    \label{tab:performance}
\end{table}

\section{Conclusion}
We study the problem of calibrating a high-dimensional mechanistic epidemic model for inference and forecasting of COVID-19 using fine-grained human mobility data. Using tools from Bayesian inference, we construct hourly contact networks capturing the interaction of individuals and the rapidly changing mobility dynamics during the pandemic. We present our data-assimilation approach for estimating the parameters of the model, AutoEKF, and we demonstrate its effectiveness by calibrating a model for the city of Philadelphia that accurately fits the observed cases and deaths both in-sample and out-of-sample. 



\bibliographystyle{IEEEtran}
\bibliography{references, gp_and_data}

\end{document}